\documentclass[11pt]{article}
 \newcommand{\Comment}[1]{{}}
 \usepackage[textwidth = 430 pt, textheight = 630 pt]{geometry}
 \usepackage{amssymb,euscript,amsmath,amsfonts}

 \usepackage{color}
 \usepackage{bbold}
 \definecolor{MyDarkBlue}{rgb}{0.15,0.15,0.45}
 \usepackage[linktocpage=true]{hyperref}
 \hypersetup{
 colorlinks=true,
 citecolor=MyDarkBlue,
 linkcolor=MyDarkBlue,
 urlcolor=MyDarkBlue,
 pdfauthor={Neil Lambert, Constantinos Papageorgakis and Maximilian Schmidt-Sommerfeld},
 pdftitle={Deconstructing (2,0) Proposals},
 pdfsubject={hep-th}
 }

 \usepackage[numbers,sort&compress]{natbib}
 \usepackage{hypernat}

 \newcommand\ignore[1]{}
 \def\one{{\,\hbox{1\kern-.8mm l}}}

 \newcommand{\tr}{\operatorname{tr}}
 
 \newcommand{\Spin}{\operatorname{Spin}}
 \def\Tr{{\rm Tr\, }}
 
 \newcommand{\SO}{\mathrm{SO}} 
 \newcommand{\SU}{\mathrm{SU}} 
  \newcommand{\pd}{\partial}
 \newcommand{\doublet}[2]{\left(\begin{array}{c}#1\\#2\end{array}\right)}

 \newcommand{\Cset}{{\,\,{{{^{_{\pmb{\mid}}}}\kern-.45em{\mathrm C}}}}}

 \newcommand{\cN}{\mathcal N}

 \newcommand{\nn}{\nonumber}
 \newcommand{\ie}{{\it i.e.~}}
 \newcommand{\eg}{{\it e.g.~}}
 
 \newcommand{\be}{\begin{equation}}
 \newcommand{\ee}{\end{equation}}
 \newcommand{\bea}{\begin{eqnarray}}
 \newcommand{\eea}{\end{eqnarray}}

\begin{document}

  \renewcommand{\thefootnote}{\fnsymbol{footnote}}

 \rightline{CERN-PH-TH/2012-339}
  \rightline{RUNHETC-2012-22}
  \rightline{LMU-ASC 84/12}

    \vspace{1.8truecm}

 \vspace{15pt}

 \centerline{\LARGE \bf {\sc Deconstructing (2,0) Proposals }}   \vspace{2truecm}  \thispagestyle{empty} \centerline{
     {\large {\bf {\sc N.~Lambert,${}^{\,a,}$}}}\footnote{On leave  from King's College  London.}$^,$\footnote{E-mail address: \href{mailto:neil.lambert@cern.ch}{\tt  neil.lambert@cern.ch}} {}
     {\large {\bf{\sc C.~Papageorgakis${}^{\,b,}$}}}\footnote{E-mail address:
                                  \href{mailto:Costis Papageorgakis  <papageorgakis@physics.rutgers.edu>}{\tt papageorgakis@physics.rutgers.edu}} {and} {\large  {\bf{\sc M.~Schmidt-Sommerfeld${}^{\,c,d,}$}}}\footnote{E-mail address:
                                  \href{mailto:Maximilian Schmidt-Sommerfeld  <m.schmidtSommerfeld@physik.uni-muenchen.de>}{\tt m.schmidtSommerfeld@physik.uni-muenchen.de} }
                                                            }

 \vspace{1cm}
 \centerline{${}^a${\it Theory Division, CERN}}
 \centerline{{\it 1211 Geneva 23, Switzerland}}
 \vspace{.6cm}
 \centerline{${}^b${\it NHETC and Department of Physics and Astronomy}}
 \centerline{{\it Rutgers University, Piscataway, NJ 08854-8019, USA}}
 \vspace{.6cm}
 \centerline{${}^c${\it Arnold-Sommerfeld-Center f\"ur Theoretische Physik, LMU M\"unchen}}
 \centerline{\it Theresienstra\ss e 37, 80333 M\"unchen, Germany}
 \vspace{.6cm}
 \centerline{${}^d${\it Excellence Cluster Universe, Technische Universit\"at M\"unchen}}
 \centerline{\it Boltzmannstra\ss e 2, 85748 Garching, Germany}

 \vspace{1.0truecm}

 \thispagestyle{empty}

 \centerline{\sc Abstract}
 \begin{center}
   \begin{minipage}[c]{380pt}{We examine the relationships between three proposals for the        six-dimensional $(2,0)$ theory: the DLCQ of \cite{Aharony:1997th,Aharony:1997an}, the        deconstruction prescription of \cite{ArkaniHamed:2001ie}, and the five-dimensional maximally        supersymmetric Yang-Mills proposal of \cite{Douglas:2010iu,Lambert:2010iw}. We show that        \cite{ArkaniHamed:2001ie} gives a deconstruction of five-dimensional maximally        supersymmetric Yang-Mills.  The proposal of       \cite{Aharony:1997th,Aharony:1997an} uses a  subset of the degrees of freedom of       five-dimensional Yang-Mills and we show that  compactification of it on a circle of finite       radius agrees with the DLCQ arising from the  proposal of \cite{Douglas:2010iu,         Lambert:2010iw} or from the  deconstruction proposal of        \cite{ArkaniHamed:2001ie}.}

 \end{minipage}
 \end{center}

 \vspace{.4truecm}

 \noindent

 \vspace{.5cm}

 \setcounter{page}{0}

 \newpage

 \renewcommand{\thefootnote}{\arabic{footnote}}
 \setcounter{footnote}{0}

 \linespread{1.1}
 \parskip 4pt

 {}~
 {}~

 \makeatletter
 \@addtoreset{equation}{section}
 \makeatother
 \renewcommand{\theequation}{\thesection.\arabic{equation}}

 \section{Introduction}

 In this paper we wish to examine the relationships between various proposals for the  six-dimensional $(2,0)$ theory. In particular, these include the Discrete Light-Cone Quantisation  (DLCQ) definition based on the instanton quantum mechanics \cite{Aharony:1997th,Aharony:1997an}, a  definition via deconstruction from a family of four-dimensional ${\cal N }=2$ superconformal field  theories based on circular quivers \cite{ArkaniHamed:2001ie} and the more recent conjecture that  the $(2,0)$ theory on an $S^1$ of radius $R_5$ is equivalent to five-dimensional maximally  supersymmetric Yang-Mills (5D MSYM) with coupling $g^2_{YM}=4\pi^2 R_5$ \cite{Douglas:2010iu,  Lambert:2010iw}. 

 The 5D MSYM conjecture relies on 5D MSYM being a consistent quantum theory at the non-perturbative  level and not just an effective theory valid below some cut-off. Without this the conjecture is  devoid of meaning since either 5D MSYM simply does not exist as a complete quantum theory, so that  the conjecture is manifestly false, or it can only be defined as the $(2,0)$ theory on $S^1$, so  that the conjecture is tautological. On the other hand this is perhaps one of the more interesting  aspects of this proposal: namely, that a perturbatively non-renormalisable and divergent  \cite{Bern:2012di} field theory is in fact non-perturbatively well-defined without additional UV  degrees of freedom. Recently, several highly non-trivial tests of this proposal have been  performed \cite{Tachikawa:2011ch,Kallen:2012cs,Hosomichi:2012ek,Kallen:2012va,Kim:2012av,  Kallen:2012zn,Gustavsson:2011ur,Bak:2012ct,Kim:2012qf}.

 From the traditional, Wilsonian, viewpoint 5D MSYM is a non-renormalisable effective theory  obtained by integrating out degrees of freedom above some scale (proportional to $1/g^2_{YM}$).  The UV is then described by a conformal field theory, which in this case is the six-dimensional  $(2,0)$ theory. The viewpoint that we explore here is different. In particular, the proposal of  \cite{Douglas:2010iu, Lambert:2010iw} is that all the states of the UV theory are already present  in 5D MSYM.  This may seem paradoxical, however the issue is that there is no physically  well-defined separation of the theory into perturbative, {\it i.e.} power series in $g^2_{YM}$,  and non-perturbative sectors. Perturbative calculations should only be viewed as low energy  approximations where the effective coupling $g_{eff} = g^2_{YM}E$ is small and hence do not probe  the UV behaviour.

 Thus we seek  other ways to define 5D MSYM. A method that comes to mind is that of deconstruction  \cite{ArkaniHamed:2001ca}. We will show how 5D  MSYM on a (discretised) circle of radius $R_4$ can  be deconstructed from an ${\cal N}=2$  superconformal circular quiver gauge theory with $N$ nodes.  In particular, for any process involving KK modes up to some finite level $L$ the correlation  functions of 5D MSYM  can be reproduced to arbitrary accuracy by taking $N$ suitably large  compared to $L$.  One could then think of the deconstruction as providing a quantum definition of  5D MSYM in terms of a well-defined theory. Indeed we will see that this relates directly the  proposal of \cite{ArkaniHamed:2001ie} to that of \cite{Douglas:2010iu, Lambert:2010iw}. In other  words, an alternative interpretation of the proposal of \cite{ArkaniHamed:2001ie} is that one  cannot deconstruct 5D MSYM on an $S^1$ of radius $R_4$ and coupling $g^2_{YM}$ without also  deconstructing the $(2,0)$ theory on a torus with radii $R_4,R_5$ where $R_5 =  g^2_{YM}/4\pi^2$,  keeping all KK modes.

 Another method to define a theory is to consider DLCQ  and we revisit the  proposal of  \cite{Aharony:1997th,Aharony:1997an}. This proposal has the miraculous feature that it only  requires knowing the dynamics of the $(2,0)$ theory on $S^1$ in the limit that $R_5\to0$. Thus it  does not require knowledge of the theory at finite $R_5 = g^2_{YM}/4\pi^2$. We will show that this  DLCQ of the $(2,0)$ theory on a circle of finite size agrees with a DLCQ obtained from 5D MSYM  defined using deconstruction or assuming it is the $(2,0)$ theory on $S^1$. 

 The rest of this paper is organised as follows. In Section \ref{deconstruction} we perform a  discretisation of one dimension in 5D MSYM and show explicitly that the resulting action is in the  same universality class as the four-dimensional deconstructed quiver theory of  \cite{ArkaniHamed:2001ie}, both leading to the action of 5D MSYM on $S^1$ in the limit where the  spacing goes to zero. Furthermore, in the spirit of deconstruction, we argue that the quantum  theory of the quiver conformal field theory can be made to be arbitrarily close to that of the  discretised 5D MSYM theory on a circle of radius $R_4$. 
 In Section \ref{section3} we review the infinite momentum frame (IMF) and DLCQ descriptions of the  $(2,0)$ theory and argue that, unlike for the DLCQ,  there is no obvious simplification of the  theory in the IMF. On the other hand we show that a reduction of the DLCQ prescription  \cite{Aharony:1997th,Aharony:1997an} of the $(2,0)$ theory on a circle of finite radius agrees  with the DLCQ description obtained from either  5D MSYM (assuming the conjecture of  \cite{Douglas:2010iu, Lambert:2010iw}) or the deconstruction proposal \cite{ArkaniHamed:2001ie}.  Finally Section \ref{sectionconclusions} contains our conclusions and further comments.

 \section{Deconstructing   5D MSYM}\label{deconstruction}

 Our aim in this section is to deconstruct 5D MSYM starting from a well-defined four-dimensional  quiver gauge theory. The idea of deconstruction is that the quiver or theory space can, in the  Higgs phase of the 4D theory, be interpreted as a discretised physical direction with spacing $a =  1/ v G$. Here $v$ is the Higgs vev and $G$ the 4D coupling. A priori, the 5D theory emerges only  at energies below $1/a$ and is UV-completed by the well-defined 4D quiver theory  \cite{ArkaniHamed:2001ca}. However, for a superconformal theory one can attempt to take the   spacing to zero, or in other words the UV cutoff to infinity. For this one needs to start with a  4D theory which does not experience a phase transition at strong coupling  \cite{ArkaniHamed:2001ie}.  We will show that the superconformal quiver gauge theory introduced in  Section~\ref{deconsetup} exactly reproduces 5D MSYM on a discretised circle by matching the two  actions. Note that our discretisation process is not quite the same as replacing the circle by a  lattice; for a discussion on how to latticise a theory while preserving some degree of  supersymmetry see \cite{Catterall:2009it}.  Rather, we will replace functions of the circle by  piecewise constant functions.  We will then proceed to discuss the relation of  \cite{ArkaniHamed:2001ie} to the proposal of \cite{Douglas:2010iu, Lambert:2010iw}.

 \subsection{Discretised 5D MSYM: Gauge Fields}

 Let us begin with the bosonic part of the  action of 5D MSYM with gauge group $\SU(K)$ 
 \be
  S_{\rm 5D}^{B} =  \frac{1}{g_{YM}^{2}}  \int d^5 x \;\Tr \Big[- \frac{1}{4}  F_{\mu\nu}  F^{\mu\nu}   - \frac{1}{2}  D_\mu X^I  D^\mu X^{I} + \frac{1}{4}[X^I,X^J][X^I, X^J]\Big]\;,
 \ee
 where $\mu,\nu = 0,...,4$ and $I,J=1,...,5$. In view of discretising and compactifying the  4-direction we will write
 \bea\label{fstrengths}
 F_{mn}&= &\pd_m A_n - \pd_n A_m - i [A_m , A_n]\cr
 F_{4m}&= &\pd_4 A_m - D_m X^6\cr
 D_4 X^I &=& \pd_4 X^I - i [X^6, X^I]\;,
 \eea
 where we have renamed $A_4 = X^6$, and  $m,n = 0,...,3$.

 In order to proceed we first discretise the line whose coordinate is $x^4$ by splitting it into an  infinite number of equal segments of length $a$ and take the fields to be constant along each  segment. This has the effect of reducing the gauge symmetries to those of four-dimensions. In the  limit that $a\to 0$ we expect that the full five-dimensional gauge symmetry is restored.
  The integral over $x^4$ becomes a Riemann sum, which  approximates the integral as $a\to 0$.  Keeping only terms which will be relevant for the gauge field $A_m$, this gives
 \be\label{actiontwo}
  S_{\rm 5D-Discr}^{\rm B- Gauge}  = \frac{ a}{g_{YM}^{2}} \int d^4 x \sum_{k=-\infty}^\infty\;\Tr  \Big[- \frac{1}{4}  F_{mn}^{(k)} F^{(k)mn} -\frac{1}{2}\tilde\pd_4 A^{(k)}_{m}\tilde\pd^4  A^{{(k)}m}\Big]\;,
 \ee
 where $\tilde\pd_4$ is a discretised version of the derivative involving the forward difference  operator
 \be
 \tilde\pd_4 f^{(k)} = \frac{f^{(k+1)}- f^{(k)} }{a}\;.
 \ee
 We then compactify the discretised direction by identifying $f^{(N+k)}\equiv f^{(k)}$ and  truncating the sum such that $N a =2 \pi R_4$
 \be
 \begin{split}
  S_{\rm 5D-Disc}^{\rm B-Gauge} = \frac{a}{g_{YM}^{2}}  \int d^4 x   \sum_{k=[\frac{N}{2}]-N+1}^{[\frac{N}{2}]}  \;\Tr  \Big[- \frac{1}{4}  F_{mn}^{(k)} F^{(k) mn}  -\frac{1}{2}\tilde \pd_4 A_{m}^{(k)} \tilde \pd^4 A^{(k)m}\Big]\;.
 \end{split}
 \ee
 As the last step we perform a discrete Fourier transform
 \be
 A_{m}^{(k)} = \frac{1}{\sqrt N}\sum_{p=[\frac{N}{2}]-N+1}^{[\frac{N}{2}]} B_{m}^{(p)} q^{k p}\ ,
 \ee
 with $q = e^{2 \pi i /N }$; Note that the reality condition on $A_m$ imposes $B^{(-k)}_m =  B^\dagger_{m(k)}$. From now on we will omit the sum ranges over Fourier mode indices, to be  understood as above.

 In terms of the Fourier modes the derivatives on the gauge fields become
 \be
 \tilde\pd_4 A_m^{(k)} = \frac{1}{\sqrt N a}\sum_{s}B_m^{(s)} q^{ks}\Big( q^s - 1\Big)  \;,
 \ee
 whereas the gauge field strengths can be organised as
 \be\label{sums}
 \begin{split}
 F_{mn}F^{mn} =(\pd^m A^n - \pd^n A^m )^2 - 2i[A_m, A_n](\pd^m A^n - \pd^n A^m )- [A_m,A_n][A^m,  A^n]\\
 = \frac{1}{N}\sum_{k,s,s'} q^{k(s - s')} (\pd_m B^{(-s)}_n - \pd_n B^{(-s)}_m )(\pd^m B^{(s')n} -  \pd^n B^{(s')m} )\\
  -  \frac{2i}{N^{3/2}} \sum_{k,s,s',s''} q^{k(s - s'-s'')}  [B^{(-s)}_m, B^{(-s')}_n] (\pd^m  B^{(s'')n} - \pd^n B^{(s'')m} )\\
  -\frac{1}{N^2}\sum_{k,s,s',s'',s'''} q^{k(s + s'-s''-s''')} [B_m^{(-s)}, B^{(-s')}_n]  [B^{(s'')m}, B^{(s''')n}]\;.
 \end{split}
 \ee

 Plugging these expressions into \eqref{actiontwo} and performing the sums over $k$ and some of the  $s$-indices, we obtain
 \be
 \begin{split}
 S_{\rm 5D-Disc}^{\rm B-Gauge} = \frac{ a}{g_{YM}^{2}} \int d^4 x \;\Tr \Big[- \frac{1}{4}\sum_s  (\pd_m B^{(-s)}_n - \pd_n B^{(-s)}_m )(\pd^m B^{(s)n} - \pd^n B^{(s)m} )\\
 +\frac{i}{2N^{1/2}}  \sum_{s,s'} [B^{(-s)}_m, B^{(-s')}_n] (\pd^m B^{(s+s')n} - \pd^n B^{(s+s')m}  ) \\
  +\frac{1}{4N}\sum_{s,s',s''} [B^{(-s)}_m, B^{(-s')}_n] [B^{(s'')m}, B^{(s+s'-s'')n}]\\
  -\frac{1}{2}\frac{1}{a^2} \sum_s | q^s-1|^2 B^{(-s)}_m B^{{(s)}m}\Big]\;.
 \label{full5dLpart1}
 \end{split}
 \ee
 In the above we used $\sum_{k=[\frac{N}{2}]-N+1}^{[\frac{N}{2}]} q^{ k (p-s)} = N \delta_{p,s}$.

 \subsection{Discretised 5D MSYM:  Scalars}

 We proceed to consider the scalar part of 5D MSYM compactified on a discretised circle. Following  the same steps as for the gauge fields and defining the Fourier transforms in terms of
 \be
   X^{(i)}_A =  \frac{1}{\sqrt{N}} \sum_s q^{is} Y^{(s)}_A\ ,
 \ee
  we arrive at
 \be
 \begin{split}
 S_{\rm 5D-Disc}^{\rm B-Scalar} = - \frac{a}{2 g_{YM}^2}\int d^4 x \Big[ \sum_s \partial_m  Y^{(s)}_A \partial^m Y^{(-s)}_A
     - \frac{2i}{\sqrt{N}} \sum_{s,s'} [ B_m^{(s)} , Y^{(s')}_A ] \partial^m Y^{(-s-s')}_A
     \\
  - \frac{1}{N} \sum_{s,s',s''} [ B_m^{(s)} , Y^{(s')}_A ] [ B^{(-s'')m} , Y^{(s''-s-s')}_A ] \Big]
   \\
  - \frac{a}{4N g_{YM}^2} \int d^4 x\sum_{s,s',s''} [ Y^{(s)}_A , Y^{(s')}_B ] [ Y^{(-s'')}_A ,  Y^{(s''-s-s')}_B ]
   \\
  + \frac{1}{g_{YM}^2} \int d^4 x\sum_s \partial_m Y^{(s)}_6 B^{(-s)m} (q^{-s}-1)\\
     - \frac{i}{\sqrt{N}g_{YM}^2} \int d^4 x \sum_{s,s'} [ B_m^{(s)} , Y^{(s')}_6 ] B^{(-s-s')m} (  q^{-s-s'} - 1 )
   \\
  + \frac{i}{\sqrt{N}g_{YM}^2} \sum_{s,s'} [ Y^{(s)}_6 , Y^{(s')}_I ] Y^{(-s-s')}_I ( q^{-s-s'} - 1  )\\ - \frac{2}{a g_{YM}^2} \int d^4 x\sum_s \sin^2\left(\frac{\pi s}{N}\right) Y^{(s)}_I  Y^{(-s)}_I
  \;,
   \label{full5dLpart2}
 \end{split}
 \ee
 where $A\in\{ I , 6 \}$. Note that there is an asymmetry between the $A=I$ and $A =6$ terms. In  particular, there is no KK mass for $Y_6$.

 \subsection{Discretised 5D MSYM: Fermions}\label{5dferms}
 The fermionic part of the 5D MSYM action is naturally given by
 \be
 S_{\rm 5D}^{F} = \frac{1}{g_{YM}^2}\int d^5 x\;\Tr\Big(-\frac{i}{2}\bar \psi_i \gamma^\mu D_\mu  \psi_i + \frac{i}{2}\bar \psi_i {\lambda^{I}_{ij}}[X_I,\psi_j]\Big)\;,
 \ee
 where  $\mu = 0,...,4$, $I = 1,...5$, $i,j = 1,...,4$. The $\psi$'s are complex 4-component  spinors of $\Spin(1,4)$ satisfying a symplectic Majorana condition and transforming in the ${\bf  4}$ of $\Spin(5)$.\footnote{For our spinor conventions we defer the reader to the Appendix.}

 However, it will be convenient to re-write this in terms of complex 2-component 4D Weyl spinors,  such that we are able to compare with the action obtained from the 4D quiver theory via the  deconstruction description. We decompose
 \bea
   \psi_1 = \begin{pmatrix} \zeta_{1} \\ -\sigma^2\zeta_{3}^* \end{pmatrix}\;,\;
   \psi_2 = \begin{pmatrix} \zeta_{2} \\ -\sigma^2 \zeta_{4}^* \end{pmatrix}\;,\;
 \psi_3 = \doublet{-i \zeta_{4}}{ -i \sigma^2\zeta_{2}^*} \;,\;
 \psi_4 = \doublet{i \zeta_{3}}{i\sigma^2\zeta_{1}^*}\;,
 \eea
 such that the symplectic Majorana condition is satisfied. Note that the action written in terms of  the $\zeta$'s will not have manifest 5D Lorentz invariance.

 The kinetic terms will give
 \be
 \begin{split}
 S_{\rm 5D}^{\rm F-Kin}=\frac{1}{g_{YM}^2}\int d^5 x\;\Tr\Big(i \bar \zeta_1 \bar \sigma^m D_m  \zeta_1 +i \bar\zeta_2 \bar\sigma^m D_m \zeta_2 +i \bar \zeta_3 \bar \sigma^m D_m \zeta_3 +i  \bar\zeta_4 \bar\sigma^m D_m \zeta_4\\ +\zeta_{3} D_4\zeta_1 -i\bar \zeta_3 D_4 \bar\zeta_1 +i  \zeta_{4} D_4\zeta_2 - i\bar \zeta_4 D_4 \bar\zeta_2  \Big)\;,
 \end{split}
 \ee
 where $\bar \zeta = \zeta^\dagger$. Use has been made of the identities $(i \sigma^2) \zeta^* =  \bar \zeta$ and $\zeta^T (i \sigma^2 ) = - \zeta$, as well as  integration by parts.

 We can also work out the Yukawa interactions. We will only explicitly write down the terms  involving $X^5$.  They are
 \bea\label{fermionX5}
 S_{\rm 5D}^{\rm F-Int} &=& \frac{1}{g_{YM}^2}\int d^5x\;\Tr\Big(-\frac{i}{2}\bar  \psi_1[X_5,\psi_1]+\frac{i}{2}\bar \psi_2[X_5,\psi_2]+\frac{i}{2}\bar  \psi_3[X_5,\psi_3]-\frac{i}{2}\bar \psi_4[X_5,\psi_4]\Big)\cr
 &=&\frac{1}{g_{YM}^2}\int d^5x\;\Tr\Big(-i[\zeta_1, \zeta_3] X_5 -i[\bar\zeta_1, \bar\zeta_3] X_5  +i[\zeta_2, \zeta_4] X_5 +i[\bar\zeta_2, \bar\zeta_4] X_5\Big)\;.
 \eea

 Similarly to the bosonic case, we can turn the 4-direction into a discretised one with spacing   $a$, so that the integral becomes a sum, the derivative becomes a forward difference operator {\it  etc.}. The discretisation procedure produces the following action for the quadratic terms that we  found above
 \be
 \begin{split}
 S^{\rm F-Kin}_{\rm 5D-Disc}=\frac{a}{g_{YM}^2}\int d^4 x \sum_{k = -\infty}^{\infty}\;\Tr\Big(
  \zeta_3^{(k)} \tilde\pd_4\zeta_1^{(k)} - i\bar \zeta_3^{(k)} \tilde\pd_4 \bar\zeta_1^{(k)} +i  \zeta_4^{(k)} \tilde\pd_4\zeta_2^{(k)} - i\bar \zeta_4^{(k)} \tilde\pd_4 \bar\zeta_2^{(k)}\\+ i  \bar \zeta^{(k)}_1 \bar \sigma^m \pd_m \zeta^{(k)}_1 +i \bar\zeta^{(k)}_2 \bar\sigma^m \pd_m  \zeta^{(k)}_2 +i \bar \zeta^{(k)}_3 \bar \sigma^m \pd_m \zeta^{(k)}_3 +i \bar\zeta^{(k)}_4  \bar\sigma^m \pd_m \zeta^{(k)}_4\\
  +   \zeta_3^{(k)} [X_6^{(k)},\zeta_1^{(k)}] - \bar \zeta_3^{(k)} [X_6^{(k)} ,\bar\zeta_1^{(k)}]
 + \zeta_4^{(k)} [X_6^{(k)},\zeta_2^{(k)}] - \bar \zeta_4^{(k)} [X_6^{(k)}, \bar\zeta_2^{(k)}]\\
 +\bar \zeta_1^{(k)} \bar \sigma^m [A^{(k)}_m, \zeta_1^{(k)}] + \bar\zeta_2^{(k)} \bar\sigma^m  [A_m^{(k)}, \zeta_2^{(k)}]
 + \bar \zeta_3^{(k)} \bar \sigma^m [A_m^{(k)}, \zeta_3^{(k)}] + \bar\zeta_4^{(k)} \bar\sigma^m  [A_m^{(k)} ,\zeta_4^{(k)}]\Big)\;.
 \end{split}
 \ee

 We then compactify the discretised direction, which truncates the sum, and also perform a discrete  Fourier transform such that
 \be
 \zeta^{(k)} = \frac{1}{\sqrt N}\sum_{p =  [\frac{N}{2}]-N+1}^{[\frac{N}{2}]}\eta^{(p)}q^{kp}\qquad\textrm{and}\qquad \bar\zeta^{(k)} =  \frac{1}{\sqrt N}\sum_{p = [\frac{N}{2}]-N+1}^{[\frac{N}{2}]}\bar\eta^{(p)}q^{-kp}\;,
 \ee
 which lets us write the $\tilde\pd_4 \zeta^{(k)}$ derivatives as
 \be
 \tilde\pd_4 \zeta^{(k)}= \frac{1}{\sqrt N a}\sum_s \eta^{(s)} q^{ks}(q^s-1)\;.
 \ee
 The final answer for the  kinetic and mass terms is
 \be\label{finalferm5d}
 \begin{split}
 S^{\rm F-Kin}_{\rm 5D-Disc}=\frac{ a}{g_{YM}^2}\int d^4 x \sum_{s}\;\Tr\Big[ -  \frac{i}{a}(1-q^{-s})\Big( \eta_3^{(s)} \eta_1^{(-s)} - \bar \eta_3^{(-s)} \bar\eta_1^{(s)}+  \eta_4^{(s)} \eta_2^{(-s)} -\bar \eta_4^{(-s)} \bar\eta_2^{(s)}\\ i\bar \eta_1^{(s)} \bar \sigma^m  \pd_m \eta_1^{(s)} + i\bar\eta_2^{(s)} \bar\sigma^m \pd_m \eta_2^{(s)} + i\bar \eta_3^{(s)} \bar  \sigma^m \pd_m \eta_3^{(s)} + i\bar\eta_4^{(s)} \bar\sigma^m \pd_m \eta_4^{(s)}\Big)\Big]\\
 +\frac{a}{g_{YM}^2\sqrt N}\int d^4 x \sum_{s,s'}\;\Tr\Big(\bar \eta_1^{(s)} \bar \sigma^m  [B^{(s-s')}_m, \eta_1^{(s')}]+\bar \eta_2^{(s)} \bar \sigma^m [B^{(s-s')}_m, \eta_2^{(s')}]\\
 +\bar \eta_3^{(s)} \bar \sigma^m [B^{(s-s')}_m, \eta_3^{(s')}]+ \bar \eta_4^{(s)} \bar \sigma^m  [B^{(s-s')}_m, \eta_4^{(s')}] \\
 +  [\eta_1^{(s)}, \eta_3^{(s')}] Y_6^{(-s-s')} - [\bar\eta_1^{(s)},\bar \eta_3^{(s')}]  Y_6^{(s+s')} \\ + [\eta_2^{(s)},\eta_4^{(s')}]Y_6^{(-s-s')} - [\bar\eta_2^{(s)},\bar  \eta_4^{(s')}] Y_6^{(s+s')}\Big)\;.
 \end{split}
 \ee

 The Yukawa interactions are dealt with in a similar way. We will once again discuss the sample  term \eqref{fermionX5}. Upon discretising  we get
 \be
 \begin{split}
   S_{\rm 5D-Disc}^{\rm F-Int}=-\frac{i a}{g_{YM^2}}\int d^4x \sum_{k =  -\infty}^\infty\;\Tr\Big([\zeta_1^{(k)}, \zeta_3^{(k)}] X^{5(k)} +[\bar\zeta_1^{(k)},  \bar\zeta_3^{(k)}] X_5^{(k)}\\ -[\zeta_2^{(k)}, \zeta_4^{(k)}] X_5^{(k)} -[\bar\zeta_2^{(k)},  \bar\zeta_4^{(k)}] X_5^{(k)}\Big)\;.
 \end{split}
 \ee
 After compactifying and Fourier transforming we end up with
 \be\label{fermion5dint}
 \begin{split} S^{\rm F-Int}_{\rm 5D-Disc}= -\frac{i a}{g^2_{YM} \sqrt N}\sum_{s,s'}\Tr\int  d^4x\Big[ ([\bar \eta_1^{(s)},\bar\eta_3^{(s')}]-[\bar \eta_2^{(s)},\bar\eta_4^{(s')}])  Y_5^{(s+s')}\\ +( [\eta_1^{(s)},\eta_3^{(s')}]-[\eta_2^{(s)},\eta_4^{(s')}]) Y_5^{(-s-s')}\Big]
 \end{split}
 \ee
 and similar expressions for other interaction terms involving different scalar components. This  concludes our discussion of 5D MSYM on a discretised  circle.

 \subsection{Deconstruction: Setup}\label{deconsetup}

 We now turn to the deconstruction picture.  As the four-dimensional starting point we will use the   $\mathcal N=2$ superconformal  $A_N$ (circular) quiver theory, in the large $N$ limit. The  expectation from \cite{ArkaniHamed:2001ie} is to obtain, upon Higgsing, a theory with enhanced  supersymmetry in 5D.

 The full action of the  $\mathcal N=2$ $A_N$ quiver theory, written in terms of $\mathcal N=1$  superfields, is given by
 \bea
 S_{\rm 4D} &=& \sum_{i=[\frac{N}{2}]-N+1}^{[\frac{N}{2}]} \tr\int d^4x \Big[\frac{1}{8 \pi  }\mathrm{Im} \Big(\tau \int d^2 \theta\; W^{\alpha(i)} W^{(i)}_\alpha\Big) - \int d^2 \theta d^2  \bar\theta\; e^{2V^{(i)}}\Phi^{\dagger(i)} e^{-2V^{(i)}}\Phi^{(i)}\cr
 &&- \int d^2 \theta d^2 \bar\theta\; e^{2V^{(i+1)}}Q^{\dagger(i)} e^{-2V^{(i)}}Q^{(i)} - \int d^2  \theta d^2 \bar\theta\; e^{-2V^{(i+1)}}\tilde Q^{(i)} e^{2V^{(i)}}\tilde Q^{\dagger(i)}\cr
 && + \int d^2 \theta\;\mathcal W^{(i)} +\int d^2 \bar \theta\;\bar{\mathcal W}^{(i)} \Big]\;,
 \eea
 with $\tau = \theta/2 \pi + 4 \pi i /G^2$, where $G$ is the four-dimensional gauge coupling. Note  that the range of the sum (\ie the labelling of the nodes of the quiver) has been conveniently  chosen so as to match the discrete mode expansion of the previous sections. With that in mind, we  will again suppress the sum ranges from now on, for brevity.

 Each node has an $\SU(K)$  gauge field and is connected to its neighbours by bifundamental and  anti-bifundamental matter fields. The trace should accordingly be thought of as being over each  term in the respective representation of $\SU(K)^{(i)}$. The superpotential encodes the matter  structure and is given by
 \be
   \mathcal W^{(i)} = -i \sqrt{2} G \tr \Big[ \tilde Q^{(i)} \Phi^{(i)} Q^{(i)} - Q^{(i)}  \Phi^{(i+1)} \tilde Q^{(i)} \Big]\;.
 \ee

 In terms of components,\footnote{We follow the conventions of \cite{AlvarezGaume:1996mv}. The  superfield expansions can be found in the Appendix.} the bosonic part of the action is then
 \be\label{bosonicfull}
 \begin{split}
   S_{4D}^{B} =  \sum_{i}  \int d^4x \;\tr \Big[- \frac{1}{4G^2}  F^{(i)}_{mn} F^{(i)mn}   -  D_m  \Phi^{(i)} D^m \Phi^{(i)\dagger}
 -  D_m Q^{(i)} D^m Q^{(i)\dagger}\\
  -  D_m \widetilde{Q}^{(i)} D^m \widetilde{Q}^{(i)\dagger} - V_S\Big]\;,
 \end{split}
 \ee
 where $m,n= 0,...,3$ and  the covariant derivatives are defined as
 \bea
   D_m \Phi^{(i)} &=& \partial_m \Phi^{(i)} - i [ A^{(i)}_m , \Phi^{(i)} ]
   \cr
   D_m Q^{(i)} &=& \partial_m Q^{(i)} - i A^{(i)}_m Q^{(i)} + i Q^{(i)} A^{(i+1)}_m
   \cr
   D_m \tilde{Q}^{(i)} &=& \partial_m \tilde{Q}^{(i)}
   - i  A^{(i+1)}_m\tilde{Q}^{(i)} + i  \tilde{Q}^{(i)} A^{(i)}_m\;.
 \eea
  The scalar potential $V_S$ is given by
 \be
 V_S = V_F + V_D\;,
 \ee
 where
 \be\label{potentials}
 V_F = \sum_i \tr \Big( F^\dagger_{Q^{(i)}}F_{Q^{(i)}} +  F^\dagger_{\tilde Q^{(i)}}F_{\tilde    Q^{(i)}}+  F^\dagger_{\Phi^{(i)}}F_{\Phi^{(i)}}\Big)\quad ,\quad V_D = \frac{G^2}{2} \sum_i  D^{(i)A} D^{(i)}_A\;,
 \ee
 with $A$ an adjoint gauge symmetry index. In turn, one has that
 \bea\label{Fterms}
 F_{Q^{(i)}} &=& -i\sqrt 2 G(\tilde   Q^{(i)} \Phi^{(i)} - \Phi^{(i+1)}\tilde   Q^{(i)})\cr
 F_{\tilde Q^{(i)}} &=& -i\sqrt 2 G( \Phi^{(i)}Q^{(i)} -  Q^{(i)} \Phi^{(i+1)} )\cr
 F_{\Phi^{(i)}} &=& -i\sqrt 2 G( Q^{(i)}\tilde   Q^{(i)} - \tilde Q^{(i-1)} Q^{(i-1)})\;,
 \eea
 for the F-terms and
 \be\label{Dterms}
 D^{(i)A} = \tr\Big[ T^A\Big( [\Phi^{(i)}, \Phi^{(i)\dagger}] + Q^{(i)}Q^{(i)\dagger} - \tilde  Q^{(i)\dagger} \tilde Q^{(i)} - Q^{(i-1)\dagger}Q^{(i-1)}+ \tilde Q^{(i-1)} \tilde  Q^{(i-1)\dagger}  \Big) \Big]\;,
 \ee
 for the D-terms.  Note that since we are working with $\SU(K)$ gauge groups,  the D-term potential   involves both single and double-trace terms coming from
 \be
 {(T^A)^i}_j {(T^A)^k}_l = \delta^i_l \delta^k_j - \frac{1}{K}\delta^{i}_j \delta^k_l\;,
 \ee
 where our normalisation for the generators is $\tr(T^A T^B) = \delta^{AB}$.

 The fermionic part of the four-dimensional theory is given in component form by the expression
 \be\label{fermionaction}
 \begin{split}
  S_{4D}^{F} = \sum_i \tr\int d^4 x\Big[ \frac{i}{G^2} \bar \lambda^{(i)} \bar\sigma^m D_m  \lambda^{(i)}+ i \bar \chi^{(i)}\bar \sigma^m D_m \chi^{(i)}
 + i \bar \psi^{(i)}\bar \sigma^m D_m \psi^{(i)}+ i \bar{\tilde{\psi}}^{(i)}\bar \sigma^m D_m  \tilde \psi^{(i)}\\
  - i \sqrt 2 (\bar \lambda^{(i+1)} \bar \psi^{(i)}- \bar \psi^{(i)} \bar \lambda^{(i)} )Q^{(i)}  -   i \sqrt 2 (\lambda^{(i)}  \psi^{(i)}-  \psi^{(i)} \lambda^{(i+1)})Q^{(i)\dagger}\\
   - i \sqrt 2 ( \lambda^{(i+1)} {\tilde{\psi}}^{(i)}- {\tilde{\psi}}^{(i)}  \lambda^{(i)} )\tilde  Q^{(i)\dagger}  -  i \sqrt 2 (\bar\lambda^{(i)}  \bar{\tilde\psi}^{(i)}-  \bar{\tilde\psi}^{(i)}  \bar{\lambda}^{(i+1)})\tilde Q^{(i)}\\
  -i\sqrt 2 G (\tilde \psi^{(i)} \chi^{(i)} - \chi^{(i+1)}\tilde \psi^{(i)})Q^{(i)} +i\sqrt 2 G  (\bar \chi^{(i)} \bar{\tilde{\psi}}^{(i)}  - \bar{\tilde{\psi}}^{(i)}  \bar\chi^{(i+1)})Q^{(i)\dagger}\\
   -i\sqrt 2 G ( \chi^{(i)}  \psi^{(i)}-  \psi^{(i)} \chi^{(i+1)})\tilde Q^{(i)} + i\sqrt 2 G (   \bar \psi^{(i)}\bar \chi^{(i)} - \bar\chi^{(i+1)} \bar \psi^{(i)} )\tilde Q^{(i)\dagger}\\
   -i\sqrt 2 G ( \psi^{(i)}  \tilde\psi^{(i)}-  \tilde\psi^{(i-1)} \psi^{(i-1)})\Phi^{(i)} + i\sqrt  2 G (  \bar{\tilde{\psi}}^{(i)} \bar\psi^{(i)}-    \bar\psi^{(i-1)}\bar{\tilde{\psi}}^{(i-1)})\Phi^{(i)\dagger}\\
   - i \sqrt 2 [\bar \lambda^{(i)}, \bar \chi^{(i)} ]\Phi^{(i)} -  i \sqrt 2 [\lambda^{(i)},   \chi^{(i)} ]\Phi^{(i)\dagger}\Big]\;.
 \end{split}
 \ee

 \subsection{Deconstruction: Gauge Fields}

 Deconstruction instructs us to expand the above theory around a real hypermultiplet vev, $\langle  Q^{(i)}\rangle  = \frac{v}{\sqrt 2} \one_{K \times K} $. This leads to a Higgsing of the gauge  group down to the diagonal subgroup $\SU(K)^N\to \SU(K)$; hence the trace (now denoted by $\Tr$)  will  be over the latter gauge group.

 Let us explicitly describe the setup of the calculation for the gauge fields. As a result of  Higgsing \eqref{bosonicfull} we get
 \be\label{actionone}
  \begin{split}
  S_{\rm 4D-Higgs}^{\rm B-Gauge} = \frac{1}{G^{2}} \sum_{i}\int d^4 x\;\Tr &\Big[- \frac{1}{4}   F^{(i)}_{mn} F^{(i)mn}\\ &    -\frac{1}{2}v^2 G^2 (2 A_m^{(i)} A^{(i)m} - A_m^{(i)} A^{(i+1)m} -  A_m^{(i+1)}A^{(i)m} ) \Big]\;.
 \end{split}
  \ee
 Note that the gauge fields have acquired a mass, but that the mass-matrix is off-diagonal
 \be
 A^{(i)} M_{ij} A^{(j)}\;,
 \ee
  where
 \be
    M =  v^2G^2  \begin{pmatrix}
     2      &     -1 &      0 &      0 &      0 &  \dots & -1 \\
     -1     &      2 &     -1 &      0 &      0 &  \dots & 0 \\
     0      &     -1 &      2 &     -1 &      0 &  \dots & 0 \\
     \vdots & \vdots & \ddots & \ddots & \ddots & \vdots & \vdots \\
     0      &  \dots &      0 &      0 &     -1 &      2 & -1 \\
     -1     &  \dots &      0 &      0 &      0 &     -1 & 2
   \end{pmatrix}\;.
 \ee
 In more compact notation the above can be expressed as
 \be
 M = v^2G^2[2 \one_{N \times N} - (\Omega + \Omega^{-1})]\;,
 \ee
  with $\Omega_{ij} = \delta_{i+1,j}$ the so-called $N\times N$ `shift' matrix. The latter can be  straightforwardly diagonalised into a `clock' matrix (see \eg  \cite{Csaki:2001em,Hashimoto:2008ij,Honma:2009bx})
 \be
 Q = \textrm{diag}(q^{[\frac{N}{2}]-N+1},...,q^{-1}, q^0,q,...,q^{[\frac{N}{2}]})\;,
 \ee
  with $q = e^{2 \pi i/N}$. To be specific,
 \be
 O^{-1} \Omega O = Q \qquad\textrm{and}\qquad O^{-1} \Omega^{-1} O = Q^{-1}\;.
 \ee
 The precise form of $O$ is given by
 \be\label{Os}
 O_{s s'} = \frac{1}{\sqrt N}q^{s s'}\qquad\textrm{and}\qquad (O^{-1})_{s s'} = \frac{1}{\sqrt  N}q^{-s s'}\;,
 \ee
 where the exponent on the RHS is a product of the two labels. Note that $O^\dagger = O^{-1}$.

 We can use this to diagonalise the mass matrices
 \be
 \tilde M = O^{-1} M O = v^2G^2[2 \one_{N \times N} -(Q +Q^{-1})]\;,
 \ee
 and the mass-matrix eigenvalues can be read off easily
 \be
 \tilde M_{kk}=v^2G^2 [2-(e^{2 \pi i k /N} + e^{-2 \pi i k /N})]= v^2G^2 \Big[2-  2\cos\Big(\frac{2\pi k}{N}\Big)\Big] = 4v^2G^2 \sin^2\Big(\frac{\pi k}{N}\Big)\;.
 \ee

 In order to implement the above at the level of the action, one needs to redefine the gauge fields   by the same matrix $O$, such that
 \be
 A^{(i)}= \frac{1}{\sqrt N }q^{ij}B^{(j)}\qquad\textrm{and}\qquad A_{(i)}= \frac{1}{\sqrt N  }q^{-ij}  B^\dagger_{(j)}\;.
 \ee
  Note  that the unitarity of $O$ and the reality of $A$ imply a reality condition for the $B$'s
 \be
 \sum_j q^{-ij} B^\dagger_{(j)}   = \sum_j q^{ij} B^{(j)}= \sum_j q^{-ij}B^{(-j)}\;,
 \ee
 where in the last step we have taken $j\to -j$ which does not affect the sum, and hence
 \be
 B^\dagger_{(j)} = B^{(-j)}\;.
 \ee

 Then for the mass term appearing in \eqref{actionone} we have
 \be
  A_{(i)} {M^i}_{j} A^{(j)}= B^\dagger_{(k)} {{{O^\dagger}^k}_i} {M^i}_{j} {O^j}_l B^{(l)} =  B^{(-k)} {\tilde M}_{k l} B^{(l)}\;,
 \ee
 while for the field strength
 \bea
 \sum_{i}F^{(i)}_{mn} F^{(i)mn} &=& \sum_{i}\Big[(\pd_m A^{(i)}_n - \pd_n A^{(i)}_m)^2  -  2i[A^{(i)}_m, A^{(i)}_n](\pd^m A^{(i)n} - \pd^n A^{(i)m} )\cr
 &&\qquad\qquad\qquad\qquad- [A^{(i)}_m,A^{(i)}_n][A^{(i)m}, A^{(i)n}]\Big]\cr
 &=& \sum_{s} (\pd_m B^{(-s)}_n - \pd_n B^{(-s)}_m) (\pd^m B^{(s)n} - \pd^n B^{(s)m})  \cr
 &&- \frac{2i}{N^{1/2}} \sum_{s,s'} [B^{(-s)}_m, B^{(-s')}_n](\pd^m B^{(s+s')n}- \pd^n B^{(s+s')m}  ) \cr
 &&-\frac{1}{N} \sum_{s,s',s''} [B^{(-s)}_m,B^{(-s')}_n][B^{(s'')m}, B^{(s+s'-s'')n}]\;.
 \eea
 In the intermediate steps of the above, one obtains sums similar to \eqref{sums}, some of which  can be explicitly performed.

 Putting everything together, we arrive at the final answer for the gauge fields
 \be\label{final4d}
 \begin{split}
 S_{\rm 4D-Higgs}^{\rm B-Gauge} =  \frac{1}{G^2}\int d^4 x\; \Tr \Big[- \frac{1}{4} \sum_{s} (\pd_m  B^{(-s)}_n - \pd_n B^{(-s)}_m) (\pd^m B^{(s)n} - \pd^n B^{(s)m})\\
  +\frac{i}{2 N^{1/2}} \sum_{s,s'}[B^{(-s)}_m, B^{-(s')}_n](\pd^m B^{(s+s')n}- \pd^n B^{(s+s')m}  )\\
  +\frac{1}{4N} \sum_{s,s',s''} [B^{(-s)}_m,B^{(-s')}_n][B^{(s'')m}, B^{(s+s'-s'')n}]\\
  - \frac{1}{2}(4v^2G^2) \sum_{s} \sin^2\Big(\frac{\pi s}{N}\Big) B^{(-s)}_m B^{(s)m} \Big]\;.
 \end{split}
 \ee

 \subsection{Deconstruction: Scalar Fields}

 We continue by considering the scalar field terms in the action. In particular we have upon  Higgsing \eqref{Fterms}
 \bea\label{Fhiggs}
 F_{Q^{(i)}} &=& -i\sqrt 2G (\tilde   Q^{(i)} \Phi^{(i)} - \Phi^{(i+1)}\tilde   Q^{(i)})\cr
 F_{\tilde Q^{(i)}} &=& -iv G( \Phi^{(i)}-  \Phi^{(i+1)} )-i\sqrt 2 G( \Phi^{(i)}Q^{(i)} -  Q^{(i)}  \Phi^{(i+1)} )\cr
 F_{\Phi^{(i)}} &=& -iv G( \tilde   Q^{(i)} -  \tilde   Q^{(i-1)}) -i\sqrt 2 G( Q^{(i)}\tilde    Q^{(i)} - \tilde Q^{(i-1)} Q^{(i-1)})\;.
 \eea
 and from \eqref{Dterms}
 \be\label{Dhiggs}
 \begin{split}
 D^{(i)A} = \Tr\Big[ T^A\Big( [\Phi^{(i)}, \Phi^{(i)\dagger}]  - \tilde Q^{(i)\dagger} \tilde  Q^{(i)} + \tilde Q^{(i-1)} \tilde Q^{(i-1)\dagger} + Q^{(i)}Q^{(i)\dagger}  -  Q^{(i-1)\dagger}Q^{(i-1)}\\
 + \frac{v}{\sqrt 2} (Q^{(i)}+ Q^{(i)\dagger}) - \frac{v}{\sqrt 2} (Q^{(i-1)}+  Q^{(i-1)\dagger})\Big) \Big]\;.
 \end{split}
 \ee
 The covariant derivatives will give
 \bea
   D_m \Phi^{(i)} &=& \partial_m \Phi^{(i)} - i [ A^{(i)}_m , \Phi^{(i)} ]
   \cr
   D_m Q^{(i)} &=& \partial_m Q^{(i)} - \frac{i}{\sqrt 2} v (A^{(i)}_m  -  A^{(i+1)}_m) - i  A^{(i)}_m Q^{(i)} + i Q^{(i)} A^{(i+1)}_m
   \cr
   D_m \tilde{Q}^{(i)} &=& \partial_m \tilde{Q}^{(i)}
   - i  A^{(i+1)}_m\tilde{Q}^{(i)} + i  \tilde{Q}^{(i)} A^{(i)}_m\;.
 \eea

 Combining the above will lead to a variety of mass and interaction terms in addition to  contributions coming from the kinetic terms. Similar to the gauge field example, the mass matrices  can be diagonalised  by working with  redefined fields
 \bea\label{redefs}
 \Phi^{(i)} &=& \frac{1}{\sqrt N}q^{ij} \hat\Phi^{(j)}  \cr
 \tilde Q^{(i)} &=&  \frac{1}{\sqrt N}q^{ij}\hat{\tilde Q}^{(j)}\cr
  Q^{(i)} &=&  \frac{1}{\sqrt N}q^{ij} \hat Q^{(j)} \;.
 \eea

 At this stage we would like to bring the reader's attention to the following fact: In the  subsequent calculation one finds that for cubic and quartic interactions involving matter fields  with different node indices there is disagreement with the discretised 5D description for generic  values of $N$. This is no cause for concern since we have already mentioned that the prescription
 of \cite{ArkaniHamed:2001ie,ArkaniHamed:2001ca} requires large $N$. In fact, in the large-$N$  limit there is a  simplification arising from the redefinitions \eqref{redefs}. Note that  in  terms of the hatted fields one has \eg
 \bea\label{largeN}
 \tilde Q^{(i)}\tilde Q^{(i-1)} = q^{ij}q^{(i-1)k}\hat{\tilde Q}^{(j)}\hat{\tilde Q}^{(k)}\simeq  q^{ij}q^{ik}\hat{\tilde Q}^{(j)}\hat{\tilde Q}^{(k)}\;,
 \eea
 for each fixed $k\ll N$. Thus, provided that we restrict attention to processes involving KK modes  up to some finite level $L$, there is no difference between $\tilde Q^{(i)}\tilde Q^{(i-1)}$ and  $\tilde Q^{(i)}\tilde Q^{(i)}$ to leading order in $N\gg L$. Hence, ignoring all $1/N$  corrections, one can write
 \bea
 F_{Q^{(i)}} &\simeq& -i\sqrt 2G [\tilde   Q^{(i)}, \Phi^{(i)}]\cr
 F_{\tilde Q^{(i)}} &\simeq& -iv G( \Phi^{(i)}-  \Phi^{(i+1)} )-i\sqrt 2 G [\Phi^{(i)},Q^{(i)}]\cr
 F_{\Phi^{(i)}} &\simeq& -iv G( \tilde   Q^{(i)} -  \tilde   Q^{(i-1)}) -i\sqrt 2 G [Q^{(i)},\tilde    Q^{(i)}]\;.
 \eea
 and
 \be
 D^{(i)A} \simeq \Tr\Big[ T^A\Big( [\Phi^{(i)}, \Phi^{(i)\dagger}] + [\tilde Q^{(i)}, \tilde  Q^{(i)\dagger}]+[ Q^{(i)}, Q^{(i)\dagger}] + \frac{v}{\sqrt 2} (Q^{(i)}+ Q^{(i)\dagger}) -  \frac{v}{\sqrt 2} (Q^{(i)}+ Q^{(i)\dagger})\Big) \Big]\;.
 \ee

 Moreover, the covariant derivatives will now be
 \bea
   D_m \Phi^{(i)} &=& \partial_m \Phi^{(i)} - i [ A^{(i)}_m , \Phi^{(i)} ]
   \cr
   D_m Q^{(i)} &\simeq& \partial_m Q^{(i)}   - i [A^{(i)}_m, Q^{(i)}]  - \frac{i}{\sqrt 2} v  (A^{(i)}_m  -  A^{(i+1)}_m)\cr
   D_m \tilde{Q}^{(i)} &\simeq& \partial_m \tilde{Q}^{(i)}
   - i  [A^{(i)}_m,\tilde{Q}^{(i)}]\;,
 \eea
 and to leading order in $1/N$, the bifundamental scalars behave as adjoints of the diagonal  $\SU(K)$.

 The above simplification also dictates that to leading order we can ignore both the trace parts of   $Q$ and $\tilde Q$ as well as the double-trace terms coming from the D-terms: First note that the  commutator structure of the F- and D-terms above is going to eliminate the trace part of the $Q$s  and $\tilde Q$s. Furthermore, any double-trace expressions coming from the second term of   ${(T^A)^i}_j {(T^A)^k}_l = \delta^i_l \delta^k_j - \frac{1}{K}\delta^{i}_j \delta^k_l$ in the  D-term potential are also going to vanish.

 With this in mind, we can treat the $\Phi$, $Q$ and $\tilde Q$ on equal footing. It will be useful  to express the complex scalars in terms of their real and imaginary parts. So we write
 \bea
 \hat \Phi^{(i)} &=& \frac{1}{G\sqrt 2}(Y^{(i)}_1- iY^{(i)}_2 )\cr
   \hat{\tilde Q}^{(i)} &=& \frac{1}{G \sqrt{2}} ( Y^{(i)}_3 - i Y^{(i)}_4 )\cr
 \hat Q^{(i)} &=& \frac{1}{G \sqrt{2}} ( Y^{(i)}_5 - i Y^{(i)}_6 )\;.
 \eea

 Now, consider terms involving only the adjoint scalars $\Phi$. The only contribution to their mass  is going to come from the F-term potential, while the quartic interaction will come from the  D-term
 \be\label{phiterms}
 \begin{split}
   S_{\rm 4D-Higgs}^{{\rm B-}\Phi} = \sum_{i}\int d^4 x\;\Tr \Big[- D_m\Phi^{(i)}D^m  \Phi^{(i)\dagger} - \frac{G^2}{2}[\Phi^{(i)},\Phi^{(i)\dagger}]^2\\   - v^2 G^2(2 \Phi^{(i)}  \Phi^{(i)\dagger} - \Phi^{(i)} \Phi^{(i+1)\dagger} - \Phi^{(i+1)}\Phi^{(i)\dagger} ) \Big]\;.
 \end{split}
  \ee
 This expression is identical in structure to the one for the gauge fields \eqref{actionone} and we  can proceed analogously. Our redefinition in terms of hatted fields diagonalises the mass-matrix  and in terms of real components one obtains, \eg for the real part of $\Phi$,
 \be
 \begin{split}
 S_{\rm 4D-Higgs}^{{\rm B-}Y_1}= \frac{1}{G^2}\int d^4 x\; \Tr \Big[- \frac{1}{2} \sum_{s} \pd_m  Y_1^{(-s)} \pd^m Y_1^{(s)}
 +\frac{i}{N^{1/2}} \sum_{s,s'}[B^{(-s)}_m, Y_1^{(-s')}]\pd^m Y_1^{(s+s')}\\
  +\frac{1}{2N} \sum_{s,s',s''} [B^{(-s)}_m,Y_1^{(-s')}][B^{(s'')m}, Y^{(s+s'-s'')}_1]\\
  +\frac{1}{4N} \sum_{s,s',s''} [Y^{(-s)}_1,Y^{(-s')}_{1}][Y^{(s'')}_1, Y^{(s+s'-s'')}_{1}]\\
  - \frac{1}{2}(4v^2G^2) \sum_{s} \sin^2\Big(\frac{\pi s}{N}\Big) Y_1^{(-s)} Y_1^{(s)} \Big]\;.
 \end{split}
 \ee

 Since at leading order in $1/N$ we can treat $\Phi$, $Q$ and $\tilde Q$ similarly, it is  straightforward to evaluate the rest of the scalar terms. The only point of special interest is  that  the field $Y_6$ does not pick up a mass during the Higgsing process and there is an  asymmetry between the $A= I$ and $A = 6$ terms. The final result is
 \be\label{finalscalars4d}
 \begin{split}
 S_{\rm 4D-Higgs}^{\rm B-Scalars} = \frac{1}{G^2}\int d^4 x\; \Tr \Big[- \frac{1}{2} \sum_{s} \pd_m  Y_A^{(-s)} \pd^m Y_A^{(s)}
  +\frac{i}{N^{1/2}} \sum_{s,s'}[B^{(-s)}_m, Y_A^{(-s')}]\pd^m Y_A^{(s+s')}\\
 +\frac{1}{2N} \sum_{s,s',s''} [B^{(-s)}_m,Y_A^{(-s')}][B^{(s'')m}, Y^{(s+s'-s'')}_A]\\
  +\frac{1}{4N} \sum_{s,s',s''} [Y^{(-s)}_A,Y^{(-s')}_{B}][Y^{(s'')}_A, Y^{(s+s'-s'')}_{B}]\\ -  \frac{1}{4N G^2} \int d^4 x\sum_{s,s',s''} [ Y^{(s)}_A , Y^{(s')}_B ] [ Y^{(-s'')}_A ,  Y^{(s''-s-s')}_B ]
   \\
  + \frac{v}{G} \int d^4 x\sum_s \partial_m Y^{(s)}_6 B^{(-s)m} (q^{-s}-1)\\
     - \frac{iv}{\sqrt{N}G} \int d^4 x \sum_{s,s'} [ B_m^{(s)} , Y^{(s')}_6 ] B^{(-s-s')m} (  q^{-s-s'} - 1 )
   \\
  + \frac{iv}{\sqrt{N}G} \sum_{s,s'} [ Y^{(s)}_6 , Y^{(s')}_I ] Y^{(-s-s')}_I ( q^{-s-s'} - 1 )
  - \frac{1}{2}(4v^2G^2) \sum_{s} \sin^2\Big(\frac{\pi s}{N}\Big) Y_I^{(-s)} Y_I^{(s)} \Big]\;.
 \end{split}
  \ee

 \subsection{Deconstruction: Fermions}

 We now proceed to study the effect of expanding the fermionic part of the action  \eqref{fermionaction} around $\langle Q^{(i)}\rangle= \frac{v}{\sqrt 2 } \one_{K \times K} $. This  gives rise to the following mass terms
 \bea
  S_{\rm 4D-Higgs}^{\rm F-mass} &=& \sum_i \Tr\int d^4 x\Big[ - i v (\bar \lambda^{(i+1)} \bar  \psi^{(i)}- \bar \psi^{(i)} \bar \lambda^{(i)} )  -  i v (\lambda^{(i)}  \psi^{(i)}-  \psi^{(i)}  \lambda^{(i+1)})\cr
 &&\qquad -iv  G (\tilde \psi^{(i)} \chi^{(i)} - \chi^{(i+1)}\tilde \psi^{(i)}) -i v G (  \bar{\tilde{\psi}}^{(i)} \bar\chi^{(i+1)}-\bar \chi^{(i)} \bar{\tilde{\psi}}^{(i)}  )\Big]\;.
 \eea
 In order to diagonalise the fermion mass matrices, define
 \be
   ( \lambda^{(i)} , \chi^{(i)} , \psi^{(i)} , \tilde \psi^{(i)} ) = \frac{1}{G\sqrt{N}}  \sum_sq^{is} (G \eta_1^{(s)} , \eta_2^{(s)} , \eta_3^{(s)} ,  \eta_4^{(s)} )\ ,
 \ee
 and note that the large-$N$ simplifications which we used in \eqref{largeN} will also apply for  products of bifundamental fermions.

 The fermion mass terms then become
 \be\label{fmass}
   S_{\rm 4D-Higgs}^{\rm F-Mass} = - \frac{i v}{G} \Tr\int d^4 x  \sum_s (1-q^{-s})   \Big[  \eta_1^{(-s)} \eta_3^{(s)} - \bar \eta_1^{(s)} \bar \eta_3^{(-s)}  +\eta_2^{(-s)} \eta_4^{(s)}  -\bar \eta_2^{(s)} \bar{\eta}_4^{(-s)} \Big]\;.
 \ee
 For the fermion kinetic terms we have
 \be\label{fkinetic}
 \begin{split}
    S_{\rm 4D-Higgs}^{\rm F-Kin} = \frac{i}{G^2} \sum_s \Tr \int d^4 x\Big[ \bar \eta_1^{(s)} \bar  \sigma^m \partial_m \eta_1^{(s)} + \bar \eta_2^{(s)} \bar \sigma^m \partial_m \eta_2^{(s)}
  + \bar \eta_3^{(s)} \bar \sigma^m \partial_m \eta_3^{(s)} + \bar{\eta}_4^{(s)} \bar \sigma^m  \partial_m \eta_4^{(s)} \Big]\\
     + \frac{1}{G^2\sqrt{N}} \sum_{s,s'} \Tr \int d^4 x\Big[ \bar\eta_1^{(s)} \bar\sigma^m [  B^{(s-s')}_m , \eta_1^{(s')} ]      + \bar\eta_2^{(s)} \bar\sigma^m [ B^{(s-s')}_m , \eta_2^{(s')}  ]\\
        + \bar\eta_3^{(s)} \bar\sigma^m ( B^{(s-s')}_m \eta_3^{(s')} - q^{s-s'} \eta_3^{(s')}  B^{(s-s')}_m )
       \\
     + \bar{\eta}_4^{(s)} \bar\sigma^m (  q^{s-s'} B^{(s-s')}_m \eta_4^{(s')} - \eta_4^{(s')}  B^{(s-s')}_m ) \Big]\;,
 \end{split}
 \ee
 while the terms involving $\Phi$ become in the large-$N$ limit
 \be
 \begin{split}
   S_{\rm 4D-Higgs}^{\rm F-Int-\Phi} = \frac{1}{G^2\sqrt{N}} \sum_{s,s'} \Tr \int d^4 x \Big[ -i   [\eta^{3(s)}, \eta^{4(s')}] ( Y_1^{(-s-s')} + i Y_2^{(-s-s')} )
 \\+ i [ \bar{\eta}_4^{(s')}, \bar \eta_3^{(s)}] ( Y_1^{(s+s')} - i Y_2^{(s+s')} )
   \\ - i [ \bar\eta_1^{(s)} , \bar \eta_2^{(s')} ] ( Y_1^{(s+s')} + i Y_2^{(s+s')} )\\
 - i [ \eta_1^{(s)} , \eta_2^{(s')} ] ( Y_1^{(-s-s')} - i Y_2^{(-s-s')} ) \Big]\;.
 \end{split}
 \ee
 Finally, for the terms involving $Q$ we will have, again in the large-$N$ limit
 \bea\label{fermionQ}
 S_{\rm 4D-Higgs}^{\rm F-Int-Q} &=& -\frac{i}{G^2 \sqrt N}\sum_{s,s'}\Tr \int d^4 x \Big[ ([\bar  \eta_1^{(s)},\bar\eta_3^{(s')}]-[\bar \eta_2^{(s)},\bar\eta_4^{(s')}]) Y_5^{(s+s')}\cr
 &&\qquad\qquad\qquad\qquad+( [\eta_1^{(s)},\eta_3^{(s')}]-[\eta_2^{(s)},\eta_4^{(s')}])  Y_5^{(-s-s')}\Big]\cr
 &&+\frac{1}{G^2 \sqrt N}\sum_{s,s'}\Tr\int d^4 x\Big[ ([\bar \eta_1^{(s)},\bar\eta_3^{(s')}]+[\bar  \eta_2^{(s)},\bar\eta_4^{(s')}]) Y_6^{(s+s')}\cr
 &&\qquad\qquad\qquad\qquad-( [\eta_1^{(s)},\eta_3^{(s')}]+[\eta_2^{(s)},\eta_4^{(s')}])  Y_6^{(-s-s')}\Big]\;.
 \eea
 and similar expressions for terms involving  $\tilde Q$.

 \subsection{Comparing Discretised 5D MSYM to Deconstruction and the (2,0) Theory}\label{comparing}

 Having obtained explicit expressions for all terms in both discretised and compactified 5D MSYM,  as well as the Higgsed $\mathcal N=2$ $A_N$ quiver theory, we are now in a position to compare the  two. We see that the kinetic and  mass terms for the  gauge field in the expressions  \eqref{full5dLpart1} and \eqref{final4d} fix the relations
 \be
 \frac{1}{G^2} = \frac{a}{g_{YM}^2}\qquad \textrm{and} \qquad G^2 v^2 = \frac{1}{a^2} \;,
 \ee
 since $|e^{2 \pi i s / N } - 1|^2 = 4\sin^2{(\frac{\pi s}{N})}$.
 These further yield
 \be\label{identification}
 a= \frac{1}{G v}\qquad \textrm{and} \qquad  g_{YM}^2 = \frac{G}{ v }\;.
 \ee
 It is straightforward to check that, with these identifications, all terms between the 5D and 4D  calculations match exactly, that is \eqref{full5dLpart2} with \eqref{finalscalars4d},  \eqref{finalferm5d} with \eqref{fmass}-\eqref{fkinetic}, and \eqref{fermion5dint} with  \eqref{fermionQ}. Thus, we arrive at the conclusion that the $A_N$ quiver theory at large $N$  deconstructs 5D MSYM on a discretised circle with spacing $a$.\footnote{Alternatively, one can  compare  the deconstructed theory to 5D MSYM compactified on a circle without prescribing any kind  of discretisation but truncating the KK tower at level $N$. This leads to a four-dimensional  action which is similar to the one we have obtained by discretisation but with two  differences.  Firstly, the mass spectrum of the fundamental fields is the familiar KK pattern of $M^2 =  L^2/R^2$. Secondly, one finds no powers of $q$ in the interaction terms.  Nevertheless the  discussion that follows is similar.
 }

 During the course of the 4D calculation, we noted (and just confirmed) that in order to get  agreement between the two descriptions we needed to ignore corrections of $\mathcal O(1/N)$. This  is not surprising. The claim of deconstruction is not that one finds 5D MSYM exactly, at all  scales. Rather, if we restrict to observables that only involve KK modes up to some level $L$ then   correlation functions of these can be computed to arbitrary accuracy in the deconstructed theory  by choosing $N \gg L$ \cite{ArkaniHamed:2001ca}. 

 Let us expand upon this point. For fixed $N$ we have two theories at hand. One is the  superconformal quiver gauge theory with $N$ nodes and coupling $G^2 = 1/v^2 a^2$. The other is the  discretised 5D MSYM on a circle of radius $R_4 = Na/2\pi$. This latter theory is a truncated  version of the full 5D MSYM on a circle, analogous to a KK reduction keeping the first $N$ levels.  It is a deformation of 4D MSYM obtained by adding a finite number of massive fields which form  complete short $\cN=4$ multiplets at each level. It is renormalisable and we view it as an  effective field theory below the scale $N/R_4$. If we examine physical processes up to some scale  $L/R_4$, we then expect the effect of any modes of 5D MSYM with energy above $N/R_4$ that we  neglected in the discretisation to be suppressed by powers of $L/N$.

 Let us now compare the quantum theories arising from these two actions. We restrict attention to  correlation functions of local operators composed from the fields that appear in the Lagrangians  but only up to some scale $L/R_4$. The correlation functions obtained in the two theories will  agree up to powers of $L/N$, arising from the differing interaction terms in the action. By taking  $N\gg L$ we can  match the computations of the discretised 5D MSYM arbitrarily well by using the  quiver theory and in particular we could take the limit $N\to \infty$ with $R_4$ fixed.   This  allows us to compute a large class of local correlators of 5D MSYM on a circle of radius $R_4$ as  a limit of the superconformal quiver gauge theories.

 The deconstruction that we have just obtained is precisely the deconstruction of the $(2,0)$  theory that was proposed in \cite{ArkaniHamed:2001ie}. These authors noted that the  four-dimensional SCFT has an $\mathrm{SL}(2,{\mathbb Z})$ S-duality which maps $G \to 2 \pi N/G$.  Hence, in addition to the perturbative spectrum of states, arising from the deconstruction of the  4-direction, with masses
 \be
 M^2_{KK1} = 4{G^2v^2}\sin^2 \left(\frac{\pi L}{N}\right) =  \frac{4}{a^2}\sin^2 \left(\frac{\pi  L}{N}\right)\qquad L\in {\mathbb Z}\;,
 \ee
 there is also a dual tower of magnetically charged soliton states with
 \be
 M^2_{KK2} = 4\frac{(2 \pi)^2  v^2N^2}{  G^2}\sin^2 \left(\frac{\pi L}{N}\right)= \frac{4 (2\pi)^2  N^2}{g_{YM}^4}\sin^2 \left(\frac{\pi L}{N}\right)\qquad L\in {\mathbb Z}\;.
 \ee
 In the limit $N\gg L$ these can be identified as two KK towers corresponding to compactification  of a six-dimensional theory with radii $R_4 = Na/2\pi$ and $R_5= g^2_{YM}/4\pi^2$, both of which  are freely adjustable parameters. In addition there will be a complete $\mathrm{SL}(2,{\mathbb  Z})$ invariant spectrum of states carrying both types of KK momenta. Thus the deconstruction  argument of \cite{ArkaniHamed:2001ie} shows that one cannot deconstruct 5D MSYM without also  simultaneously deconstructing a six-dimensional theory with 16 supersymmetries and an $\SO(5)$  R-symmetry -- presumably the $(2,0)$ theory. 

 In the limit $a\to 0$, $N\to \infty$, with $R_4\to\infty$   the 4D theory results are matched to  uncompactified 5D MSYM with arbitrary value for the coupling $g_{YM}^2$, which suggests that  deconstruction in principle provides a quantum definition of 5D MSYM at all scales. Finally, the  remaining tower of KK modes of masses $M_{KK2}$ is nothing but the instanton-soliton tower of  \cite{Rozali:1997cb,Berkooz:1997cq}.  This matches the content of the conjecture of  \cite{Douglas:2010iu, Lambert:2010iw}.

 We note that since 5D MSYM is not well-defined (at least naively) we cannot claim that the  deconstructed theory is 5D MSYM. Rather, our discussion shows that deconstruction provides a  controlled definition of 5D MSYM.   Our purpose here was to find a way of identifying parameters  between the 4D and 5D descriptions and this approach has enabled us to do so in a natural way. 

 \section{The DLCQ of the (2,0) theory and 5D MSYM}\label{section3}

 We now shift gears and turn our attention to the (2,0) proposals of  \cite{Aharony:1997th,Aharony:1997an}. In order to compare the latter to the proposal that the  $(2,0)$ theory on $S^1$ is equivalent to 5D MSYM, we will need to quickly review the philosophy  behind the Infinite Momentum Frame (IMF) and the related Discrete Light-Cone Quantisation (DLCQ).  There are various outstanding conceptual and technical issues with the IMF, and especially DLCQ,  which need to be addressed before one can claim to have a complete understanding of a theory that  is defined using these methods. However, it is not our intention to resolve or discuss these  issues here. Rather, we will accept the IMF and DLCQ prescriptions at face value and focus on the  arguments leading to the proposal  of \cite{Aharony:1997th,Aharony:1997an}.

 \subsection{IMF and DLCQ}

 The basis for the IMF is that, since we are considering a Lorentz-invariant field theory, we can  examine it in any frame we like. By a judicious choice of frame the physics might be simpler to  analyse. To this end let us consider  an M5-brane wrapped on an $S^1$ of radius $R_5$ and boost it  along the compact $x^5$ direction. The energies and momenta transform as ($E=P_0$)
 \bea\label{boosted}
 E'  &=& \frac{1}{\sqrt{1-u^2}}(E-uP_5)\nonumber\\
 P_5' &=& \frac{1}{\sqrt{1-u^2}}(P_5-uE)\nonumber\\
 P'_i &=& P_i\;,
 \eea
 where $i=1,...,4$. If we introduce light-cone coordinates
 $$
 x^\pm = \frac{1}{\sqrt 2} (t\pm x^5)\;,\qquad P_\pm = \frac{1}{\sqrt 2} (E\pm  P_5)\ ,
 $$
 then \eqref{boosted} can be written as
 \bea
 P_+' &=& \sqrt{\frac{1-u}{1+u}}P_+ \nonumber\\
 P_-'  &=& \sqrt{\frac{1+u} {1-u}}P_-\nonumber\\
 P'_i &=& P_i\;.
 \eea
 Let us write $u=(1- \epsilon^2)/(1+\epsilon^2) $  so that an infinite boost corresponds to  $\epsilon\to 0$. This limit defines the IMF.  In what follows we always only consider the term of  leading order  in $\epsilon$. We find that to leading order \eqref{boosted} becomes
 \bea\label{imf}
 P_+' &=& \epsilon P_+\nonumber\\
 P_-'  &=& \frac{1}{\epsilon}P_-  \nonumber\\
 P'_i &=& P_i\;.
 \eea
 Thus if we view the original $S^1$ as an orbifold:
 \be
 ( t , x^i , x^5 ) \cong
 ( t , x^i , x^5 + 2\pi R_5 ) ,
 \ee
 then in the IMF we have
 \be\label{nullorbifold1}
 ( {x^+}' , {x^-}' , {x^i}' ) \cong
 ( {x^+}' +  2\pi  R_+ , {x^-}' , {x^i}' ),
 \ee
 where
 \be
 R_+ =R_5/\sqrt{2}\epsilon \ .
 \ee

 Next let us consider on-shell modes in the un-boosted frame with momentum  $P_5= n/R_5$ for some  integer $n$. These have energy
 \bea
 E &=& \sqrt{P_5^2+P_i^2+  m^2}\nonumber\\
 &=& \frac{|n|}{R_5}\left(1 + F\left(\frac{R_5^2P_\perp^2 }{n^2}\right)\right)\;,
 \eea
 where we have denoted $P^2_\perp = P_i^2+m^2$ and $F(x) = \sqrt{1+x}-1= \frac{1}{2}x+\ldots$. Here  $m$ allows for the possibility of massive states that can arise on the Coulomb branch. We see that
  \bea
 P_+' &=& \frac{|n|+n}{\sqrt{2}  R_5}\epsilon+  \frac{|n|\epsilon}{\sqrt{2} R_5}F\left(\frac{   R_5^2P_\perp^2}{n^2}\right) \nonumber\\
  P_-'  &=& \frac{|n|-n}{\sqrt{2}\epsilon R_5}+  \frac{|n| }{\sqrt{2}  R_5\epsilon}F\left(\frac{R_5^2P_\perp^2}{n^2}\right) \nonumber\\
 P'_i &=& P_i\;.
  \eea
 We find that modes with $n<0$ have diverging $P'_-$ in the IMF and thus decouple. Therefore we can  simply  look at the effective theory with these modes integrated out. This can be made arbitrarily  precise by taking $\epsilon$ suitably small. Therefore we restrict to $n>0$ for which
   \bea\label{PpreIMF}
 P_+' &=& \frac{ \sqrt{2}n\epsilon}{  R_5}+  \frac{n\epsilon}{\sqrt{2} R_5}F\left(\frac{   R_5^2P_\perp^2}{n^2}\right) \nonumber\\
  P_-'  &=&  \frac{n }{\sqrt{2} R_5\epsilon}F\left(\frac{R_5^2P_\perp^2}{n^2}\right)  \nonumber\\
 P'_i &=& P_i\;.
  \eea

 Now in the original theory we have states with all values of $n$. However  for fixed $R_5$, in the  $\epsilon\to 0 $ limit the finite momentum  states are those that have large $n$, with   $n\epsilon$ finite.  This is the traditional IMF picture (as used \eg  in \cite{Banks:1996vh}) and  is valid for any finite $R_5$ but  takes $n\to\infty$. Physically this corresponds to the fact  that the only modes left in the  infinite momentum frame are those that were moving sufficiently  fast against the boost so that  they have finite velocity after the boost.  Note also that for any  given $\epsilon$ there are  still infinitely many $n$ that must be included if one wishes to  describe the full theory.

 There is another possibility which is to take $R_5$ small with $R_+=R_5/\sqrt{2}\epsilon$ fixed.    This is the DLCQ construction and does not require  large $n$.\footnote{See \eg  \cite{Susskind:1997cw,Seiberg:1997ad} for DLCQ in the context of Matrix Theory.} Fixing $n$ here  simply means truncating to a fixed momentum sector of the theory. One must then still allow $n$ to  be arbitrary in order to describe the full theory.

 In either case we find
    \bea\label{PIMF}
 P_+' &=& \frac{ \sqrt{2}n\epsilon}{  R_5} \nonumber\\
  P_-'  &=&  -\frac{R_5 }{2\sqrt{2}n\epsilon}(P_i^2+ m^2) \nonumber\\
 P'_i &=& P_i\;.
  \eea
 Note that we have the three parameters $R_5,n$ and $\epsilon$ and in the limit $\epsilon\to0$ we  have just one constraint, namely that $n\epsilon /R_5$ is fixed.  To arrive at (\ref{PIMF}) from  (\ref{PpreIMF}) we simply require that $ P_\perp R_5/n \ll 1$ in the limit $\epsilon \to 0$.

  \subsection{Application to the (2,0) theory}\label{3.2}

  The DLCQ construction of \cite{Aharony:1997th,Aharony:1997an} works as follows. In the limit that   $R_5$ is small the $(2,0)$ theory on $S^1$ is well described by weakly coupled 5D MSYM with   coupling $g^2_{YM} = 4\pi^2 R_5$. As observed in \cite{Aharony:1997th,Aharony:1997an}, this is   something of a miracle since lagrangian field theories usually  become strongly coupled when  compactified on a small circle. In this limit states with $n$ units of momentum along the compact  direction correspond to solitons that carry  instanton number $n$. These states are heavy when  $R_5$ is small so that keeping $P_\perp$ fixed  means slow motion in the transverse directions.  This is the Manton approximation  \cite{Manton:1981mp} for solitons whereby the relevant degrees  of freedom correspond to motion on  the soliton moduli space.  In the limit that $g^2_{YM}\sim  R_5\to0$ all other interactions can be  neglected. Thus we find that the theory reduces to motion  on the moduli space of $n$ instantons.\footnote{We will come back to the details of this  derivation shortly.}  In the IMF, this corresponds to the second possibility discussed at the  end  of the previous section: For fixed $R_+ = R_5/\sqrt 2 \epsilon$, sending $R_5\to 0$ also  requires  $\epsilon\to 0$ and therefore the DLCQ description of the $(2,0)$ theory is given by  quantum  mechanics on the $n$-instanton moduli space.

 There also exists an alternative derivation directly  from  the $(2,0)$ system of  \cite{Lambert:2010wm}. This system is essentially  5D MSYM covariantly embedded into 6 dimensions  using a non-dynamical vector field $C^\mu$. Choosing $C^\mu$ spacelike, $C^\mu = g^2_{YM}\delta  ^\mu_5$, leads to 5D MSYM along $x^0,...,x^4$ with coupling $g^2_{YM}$. However one can also  consider a null embedding corresponding to an infinitely boosted D4-brane. Deferring to  \cite{Lambert:2011gb} for the details, we simply wish to observe here that for the choice $C^\mu=  g^2\delta^\mu_+$, where $g^2$ is an arbitrary parameter with dimensions of length, one finds
 \bea\label{P20}
 {\cal  P}_+   &=& -\frac{4\pi^2n}{g^2} \nonumber\\
 {\cal   P}_- &=& \frac{1}{2g^2} g_{\alpha\beta} \partial_-m^\alpha \partial_-m^\beta \nonumber\\
 {\cal  P}_i  &=&  \frac{1}{2g^2}{\rm Tr}\int d^4x F_{ij}F_-{}^j\;,
 \eea
 where $n$ is the instanton number, $g_{\alpha\beta}$ is the metric on the $n$-instanton moduli  space with coordinates $m^\alpha$ and $F_{ij}$, $F_{i-}$ are obtained from the field strength of  the instanton and are determined by the ADHM construction.\footnote{There is also a generalisation  that arises on the Coulomb branch, where one finds that ${\cal   P}_-  =\frac{1}{2g^2}\left(g_{\alpha\beta}(\partial_-m^\alpha-L^\alpha)(\partial_-m^\beta-L^\beta)+V\right)$.  Here $V\propto g_{\alpha\beta}L^\alpha L^\beta$ and $L^\alpha$ is a tri-holomorphic Killing vector  on the instanton moduli space  \cite{Lambert:2011gb}.}

 We note that the derivation of (\ref{P20}) in \cite{Lambert:2011gb} is exact, starting from the   the $(2,0)$ system of \cite{Lambert:2010wm}, and does not require taking the limit of an infinite  boost.  Examining the spectrum of ${\cal  P}_+$ shows that it can be identified with that of the  $(2,0)$ theory reduced on a null circle obtained by the identification
 \be\label{nullorbifold2}
 x^+ \cong x^+ + \frac{g^2}{2\pi}\;.
 \ee
 Comparing (\ref{P20}) with (\ref{PIMF}) we find that ${\cal P}_+ = P_+$ if we identify  $ R_+=  R_5/\sqrt{2}{\epsilon} = g^2/4\pi^2$. In addition (\ref{nullorbifold2}) precisely matches  (\ref{nullorbifold1}). Thus for any finite value of $g^2$, we obtain the DLCQ picture of the  $(2,0)$ theory. In particular, finite $g^2$ requires that $R_5=g^2_{YM}/4\pi^2\to 0$ as  $\epsilon\to 0$ and so again we only need the extreme IR of 5D MSYM.

 Having obtained the DLCQ at finite $R_+$ it  would   appear that we can arrange for $R_+\to\infty$   in the limit that $R_5,\epsilon \to 0$, leading to a description of the uncompactified  $(2,0)$  theory. At this stage one needs to be careful: In a null compactification the radius of the  ${x^+}$ identification is not physically meaningful by itself as a Lorentz boost can rescale  $R_+$. But in the IMF, when we choose a specific frame with fixed $P_+ = n/R_+$, this can be done  if we  also scale $n\to \infty$ \cite{Seiberg:1997ad}. This gives the DLCQ description of the  $(2,0)$ theory on ${\mathbb R}^{1,5}$ as the large-$n$ limit of super-quantum mechanics on the  instanton moduli space \cite{Aharony:1997th,Aharony:1997an}.

 This conclusion is a puzzling feature of the DLCQ proposal, since we have somehow managed to  define the six-dimensional $(2,0)$ theory, which is the UV of 5D MSYM, by only using information  contained in the extreme IR of 5D MSYM. The origin of this is the miracle mentioned above, namely  that the $(2,0)$ theory becomes weakly coupled when compactified on a circle with a small radius.  This miracle is also important for S-duality of four-dimensional maximally supersymmetric  Yang-Mills since it ensures that the four-dimensional coupling constant is given by the ratio of  the two circle radii and hence that modular transformations, such as interchanging the two  circles, map strong to weak coupling \cite{Witten:2009at}.

 In addition, one needs to keep in mind that, on top of the usual concerns about the IMF and DLCQ,  the instanton moduli space has singularities which must somehow be dealt with in the quantum  theory. For example the authors of \cite{Aharony:1997th,Aharony:1997an} provide one resolution in  terms of turning on mild  non-commutativity as a regulator,\footnote{A non-commutative deformation  corresponds to blowing up the moduli space singularities.} in view of switching it back off at the  end of any explicit calculation.

  \subsection{The $(2,0)$ theory in the IMF}

 Having reviewed the DLCQ proposal, let us instead consider an IMF prescription.  If we assume that  the $(2,0)$ theory on $S^1$ is equivalent to 5D MSYM, then we can also consider a traditional IMF  description of the $(2,0)$ theory with a finite value for $R_5>0$ but large $n$. 

 First, what is  5D MSYM at large $n$ and low energies\footnote{Here, low energies means small  excitation energy above the BPS bound of the $n$-instanton sector.} described by?  It is given by  an expansion around the ground state in the $n$-th soliton sector. At low velocities this leads to  the Manton approximation of slow motion on the soliton moduli space \cite{Manton:1981mp}. For a  recent detailed discussion on instanton-solitons of 5D MSYM see \cite{Allen:2012rp}.

 However, now we would like to see what happens to this description beyond the low-energy  approximation. It is well known that  the Manton approximation is not   exact,   although in the  case of monopoles in four dimensions it can be shown to be valid as long as the moduli velocities  remain small \cite{Stuart:1994tc}. Nevertheless at finite $g^2_{YM}$ the Manton approximation   does not capture all the dynamics of 5D MSYM. For the particular example of monopoles in four  dimensions there are estimates that the radiation produced in soliton scattering scales as the  third or fifth power of the velocity \cite{Manton:1988bn,Irwin:2000rc}. We expect that the  instanton-soliton solutions relevant for 5D MSYM will behave in a similar way: At non-zero  velocity these effects can only be neglected in the strict $g^2_{YM}\to 0$ limit. Therefore the  IMF picture does not simply reduce to quantum mechanics on the instanton moduli space for finite  $R_5 = g^{2}_{YM}/4\pi^2$. Rather, it contains an infinite number of additional radiation modes  that represent fluctuations about the soliton.

 One might hope that in the large-$n$ limit there could be a further suppression of the massive   modes so that the Manton approximation is again valid. For example, the solitons become heavy at  large $n$ so their centre-of-mass velocity must be bounded by $1/n$ to ensure that the momentum  remains small. However, this seems unlikely to extend to all moduli as the large-$n$ moduli space  contains configurations where the solitons are widely separated. It would then seem that various  `light' massive modes, obtained in the small-$n$ moduli space, can be lifted to the large-$n$  moduli space. For instance, one can imagine configurations of well separated solitons where only a  few are moving, in which case their velocities are not required to be small to ensure that the  total excitation energy of the system is small. Therefore, the radiation and other non-zero modes  seem to be insensitive to the value of $n$ and we do not expect any additional suppression at  large $n$. 

 Hence there does not appear to be any significant simplification by considering the $(2,0)$ in the  IMF as there was with DLCQ.

 \subsection{A DLCQ of 5D MSYM}\label{dlcqsection}

 To complete the circle of ideas we can also consider a DLCQ of 5D MSYM and compare it to a  compactified version of the proposal \cite{Aharony:1997th,Aharony:1997an}. To this end, let us  start with 5D MSYM and compactify on $x^4$ with radius $R_4$. To construct a DLCQ it is sufficient  to only consider a small $R_4$. Compactifying 5D MSYM on $S^1$ with coupling $g^2_{YM}$ we find 4D  MSYM with coupling $G^2 = g^2_{YM}/2\pi R_4$, coupled to a tower of KK modes.  For small $R_4$  this is strongly coupled but if the  proposal of \cite{Douglas:2010iu, Lambert:2010iw} is true  then 5D MSYM on $S^1$ admits an S-duality since it is the $(2,0)$ theory on $S^1\times S^1$.  Evidence for this can be found in \cite{Tachikawa:2011ch}. Alternatively, we could use the 4D  quiver deconstruction approach  as the quantum definition of 5D MSYM, as argued in  Section~\ref{comparing}. This has a built-in S-duality for the theory on a (discretised) circle.

 Applying  S-duality we arrive at weakly coupled 4D Yang-Mills with gauge coupling $G^2 \propto  2\pi R_4/g^2_{YM}\to 0$   but where the tower of KK modes around $x^4$ are now given by their  S-duals which will be some sort of monopole states. What exactly are these? From the point of view  of the $(2,0)$ theory S-duality corresponds to swapping $x^4$ with $x^5$. Therefore S-duality  takes momentum modes around $x^4$ to momentum modes around $x^5$. In 5D MSYM momentum modes around  $x^5$ are given by instanton-solitons along the ${\mathbb R}^4$  spanned by $x^1,...,x^4$.  Compactifying  this ${\mathbb R}^4$ to ${\mathbb R}^3\times S^1$ leads to  so-called calorons,  namely instantons that are periodic along $x^4$ (monopoles are special cases of calorons that are  invariant along $S^1$). As before we decompactify the theory by taking $n,R_4/\epsilon\to \infty$.   Thus we find that the DLCQ of 5D MSYM on ${\mathbb R}^{1,4}$ is given by quantum mechanics of the  caloron moduli space (see also \cite{Ganor:1997jx} for a related discussion).  Note that the use  of S-duality was crucial for this argument. As we mentioned this is guaranteed in the
 deconstructed theory. If we start with 5D MSYM on its own then S-duality on an $S^1$ of finite  size is tantamount to assuming that it is the $(2,0)$ theory on $S^1\times S^1$ and therefore  essentially assumes the conjecture of \cite{Douglas:2010iu, Lambert:2010iw}.

 Let us compare this with the DLCQ proposal constructed above for the $(2,0)$ theory on ${\mathbb    R}^{1,5}$. To obtain a DLCQ of the $(2,0)$ theory on ${\mathbb R}^{1,4}\times S^1$ we can take an  orbifold that acts as $x^5\cong x^5 + 2\pi R_5$ and hence need to consider instantons that are  periodic: once again one is lead to the moduli space of calorons. Thus we find agreement between  the DLCQ of the $(2,0)$ theory on ${\mathbb R}^{1,4}\times S^1$ at finite radius and the DLCQ of  5D MSYM on ${\mathbb R}^{1,4}$ at finite coupling. 

 \section{Conclusions}\label{sectionconclusions}

 In this paper we have discussed the relationships between  three proposals for the $(2,0)$ theory:  deconstruction \cite{ArkaniHamed:2001ie}, DLCQ \cite{Aharony:1997th,Aharony:1997an} and 5D MSYM  \cite{Douglas:2010iu, Lambert:2010iw}. In particular  we explicitly showed how   deconstruction  leads to the action of 5D MSYM. This provides a definition of the full 5D MSYM  as a limit of a  family of well-defined four-dimensional superconformal field theories.  Furthermore we showed how  the DLCQ construction is also consistent with the view that the $(2,0)$ theory on $S^1$ is given  by 5D MSYM by showing that they both lead to the same DLCQ of the $(2,0)$ theory on a finite  circle. This crucially assumed the S-duality property of the 5D theory, which is explicit when  defined in terms of deconstruction.

 A common feature of all these proposals is that they do not require any new states that do not    appear in 5D MSYM to describe the $(2,0)$ theory. This is the central observation in  \cite{Douglas:2010iu,Lambert:2010iw} and therefore it is compatible with both  \cite{Aharony:1997th,Aharony:1997an} and \cite{ArkaniHamed:2001ie}. Conversely, these proposals   provide some support to the claim  that  although 5D MSYM is perturbatively divergent  \cite{Bern:2012di} and power counting non-renormalisable, it should not simply be viewed as the  low-energy effective theory of the $(2,0)$ theory on $S^1$ in the Wilsonian sense, meaning that  some heavy states have been integrated out. Rather, the spectrum and interactions are those of the  $(2,0) $ theory. The proposals of \cite{Aharony:1997th,Aharony:1997an} and  \cite{ArkaniHamed:2001ie} offer alternative methods for computing physical quantities beyond the  techniques of traditional perturbative quantum field theory.

 In addition, there are still other ways that may be used to define the $(2,0)$ theory starting  from the conformal field theory of an arbitrary number of M2-branes  \cite{Aharony:2008ug}. For  example one can consider a large number of M2-branes that are blown-up via a Myers effect into  M5-branes wrapped on an $S^3$ of finite radius. It was argued in \cite{Lambert:2011eg} that the  resulting fluctuations of the M5-branes are given by 5D MSYM  on $S^2$, where $S^3$ is viewed as a  Hopf fibration over $S^2$. Furthermore 5D MSYM on a three-torus of finite size  can be obtained  from cubic arrays of M2-branes \cite{Jeon:2012fn}. In principle all these proposals give  definitions of 5D MSYM and the $(2,0)$ theory on $S^1$. It remains to be seen if they are  equivalent.

 Finally we should also mention some other recent proposals for the $(2,0) $ theory which we did  not discuss. One very interesting proposal is \cite{Kim:2012tr} which  considers the $(2,0)$  theory on ${\mathbb R}\times S^5$. Realising $S^5$ as a Hopf bundle over ${\mathbb CP^2}$ one can  then perform an  ${\mathbb Z}_k$ orbifold that acts on $U(1)$ fibre. In the large $k$ limit one  therefore finds the $(2,0)$ theory is given by 5D MSYM on ${\mathbb R}\times {\mathbb CP}^2$. In  this scenario the  radius of  ${\mathbb CP}^2$ determines the scale $g^2_{YM}$ and $1/k$ plays the  role of a dimensionless coupling constant. Furthermore, since $k$ is discrete, there is hope that  perturbation theory about large $k$ is finite, allowing one to extrapolate to small $k$.

There have been other recent conjectures for the $(2,0)$ theory which focus on novel lagrangian descriptions for the self-dual tensor directly in six-dimensions \cite{Chu:2012um} or five-dimensional models which include the KK towers of states \cite{Ho:2011ni,Bonetti:2012st}. It would be interesting to understand the relation of these papers to the conjectures we discuss here. In particular, since the proposals discussed here capture at least a significant portion of the dynamics of the $(2,0)$ theory, these other proposals should be related to them in some concrete way.

\section*{Acknowledgements}

We would like to thank T.~Banks, J.~Drummond, C.~Grojean, A.~Karch, G.~Moore, A.~Royston, D.~Shih and especially O.~Aharony  for discussions and comments. CP is supported by the U.S. Department of Energy under grant DE-FG02-96ER40959. MSS is supported by an EURYI award of the European Science Foundation.

\section*{Appendix}\label{appendix}

Here we collect our conventions for the gamma-matrices used in Section~\ref{5dferms}.
The $\mathrm{Spin}(1,4)$ gamma-matrices are given by
\be
  \gamma^\mu = \Bigg\{ \begin{pmatrix} 0 &  i\sigma^m \\ i \bar \sigma^m & 0 \end{pmatrix} ,
      \begin{pmatrix} -\one_{2\times 2 } & 0 \\ 0 &\one_{2\times 2 } \end{pmatrix} \Bigg\} \;,\;
  C_5 = \begin{pmatrix} \sigma^2 & 0 \\ 0 & \sigma^2 \end{pmatrix}\;,
\ee
with $m = 0,...,3$ and  $\sigma^m = \{\one, \sigma^i\}$ and $\bar\sigma^m = \{\one, -\sigma^i\}$. They satisfy
\be
\begin{split}
  \{\gamma^\mu,\gamma^\nu\} = 2 \eta^{\mu\nu}\;\qquad  \gamma^0 (\gamma^\mu)^\dagger \gamma^0 =\gamma^\mu\;\qquad \gamma^0 \gamma^1 \gamma^2 \gamma^3 \gamma^4 = -i \\
C_5 \gamma^\mu C_5^{-1} = (\gamma^\mu)^T \qquad
  (C_5 \gamma^\mu)^T = - C_5 \gamma^\mu\ \qquad
  C_5^T = -C_5 \;.
\end{split}
\ee
On the other hand, the hermitian $\mathrm{Spin}(5)$ gamma-matrices are given by
\begin{eqnarray}
\lambda^1  = \left( \begin{array}{cc}
					\sigma^2 & 0 \\
					 0 & -\sigma^2 \end{array} \right) ,\;
\lambda^2 = \left( \begin{array}{cc}
					-\sigma^1 & 0\\
					0 & \sigma^1\end{array} \right) ,\;
\lambda^3 = \left( \begin{array}{cc}
					0 & \one_{2 \times 2} \\
					 \one_{2 \times 2}  & 0\end{array} \right) , \cr
\lambda^4 =\left( \begin{array}{cc}
					0 & - i \one_{2 \times 2} \\
					 i \one_{2 \times 2}  & 0\end{array} \right) ,\;
\lambda^{5} =\left( \begin{array}{cc}
					-\sigma^3 & 0 \\
					0 & \sigma^3 \end{array} \right) ,\;
 K = \left( \begin{array}{cc}
					0  & i \sigma^2 \\
					i \sigma^2 & 0 \end{array} \right)\ ,
\end{eqnarray}
and satisfy
\be
\begin{split}
  \{ \lambda^I,\lambda^J \} = 2 \delta^{IJ}\;, \quad
  (\lambda^I)^\dagger = \lambda^I\;, \quad
  \lambda^1 \lambda^2 \lambda^3 \lambda^4 \lambda^{5} = 1\\
K^T = -K\;, \quad\; (K \lambda^I)^T = - K \lambda^I  \;.
\end{split}
\ee
The conjugate fermions are defined as
\be
\bar \psi_i \equiv \psi^{\dagger}_i \gamma^0\ ,
\ee
and satisfy the symplectic Majorana condition
\be\label{reality5d}
\bar \psi_i \ = \psi_j^T C_5 {K^j}_i\;.
\ee

We also give the superfield expansion, used in Section~\ref{deconsetup}. In Wess-Zumino gauge one has that
\bea
V^{(i)} &=& - \theta \sigma^m \bar \theta A_m^{(i)} + i \theta^2 \bar \theta \bar \lambda^{(i)} - i \bar \theta^2 \theta \lambda^{(i)} +\frac{1}{2}\theta^2 \bar \theta^2 D^{(i)}\cr
W^{(i)}_\alpha &=& - i \lambda^{(i)}_\alpha + \theta_\alpha D^{(i)} -\frac{i}{2}(\sigma^m \bar \sigma^n \theta)_\alpha F^{(i)}_{mn}+ \theta^2 (\sigma^m D_m \bar \lambda^{(i)})_\alpha  \cr
\Phi^{(i)} &=& \Phi^{(i)} + i \theta \sigma^m\bar \theta \pd_m\Phi^{(i)}-\frac{1}{4}\theta^2 \bar \theta^2\Box \Phi^{(i)}+ \sqrt 2 \theta \chi^{(i)}-\frac{i}{\sqrt 2}\theta^2 \pd_m \chi^{(i)}\sigma^m \bar \theta+ \theta^2 F_{\Phi^{(i)}}\ ,\nn\\
\eea
for the vector multiplets while
\bea
Q^{(i)} &=& Q^{(i)} + i \theta \sigma^m \bar \theta \pd_m Q^{(i)}-\frac{1}{4}\theta^2 \bar \theta^2\Box Q^{(i)}+ \sqrt 2 \theta \psi^{(i)}-\frac{i}{\sqrt 2}\theta^2 \pd_m \psi^{(i)}\sigma^m \bar \theta+ \theta^2 F_{Q^{(i)}}\cr
\tilde Q^{(i)\dagger} &=& \tilde Q^{(i)\dagger} - i \theta \sigma^m \bar \theta \pd_m \tilde Q^{(i)\dagger}-\frac{1}{4}\theta^2 \bar \theta^2\Box \tilde Q^{(i)\dagger}+ \sqrt 2 \bar \theta \bar{\tilde{\psi}}^{(i)}+\frac{i}{\sqrt 2}\bar \theta^2 \theta \sigma^m \pd_m \bar{\tilde{\psi}}^{(i)} + \bar \theta^2 F^\dagger_{\tilde Q^{(i)}}\ ,\nn\\
\eea
for the hypermultiplets, where we are using the same letter for the chiral superfields as for their scalar components in the hope that no confusion will arise.

\bibliographystyle{utphys}
\bibliography{deconstruction}

\end{document}